\newcommand{\demodsensPHA}{1.1\pm0.1 \, \mu\mathrm{rad}/\sqrt{\mathrm{Hz}}}
\newcommand{\demodsensONE}{4.0\pm0.3 \, \mathrm{mK}/\sqrt{\mathrm{Hz}}}
\newcommand{\demodsensTHREE}{11\pm 1 \, \mathrm{mK}/\sqrt{\mathrm{Hz}}}

\newcommand{\ioleverarm}{0.74 \pm 0.02}
\newcommand{\iotunnel}{270 \pm 20 \, \mathrm{ns^{-1}}}
\newcommand{\ioA}{5.13 \pm 0.06 \, \mathrm{rad\, pF^{-1}}}
\newcommand{\ioisd}{6.9 \pm 0.5\, \mathrm{nA}}
\newcommand{\bfleverarm}{0.84 \pm 0.03}
\newcommand{\bftunnel}{510 \pm 10 \, \mathrm{ns^{-1}}}
\newcommand{\bfA}{0.75 \pm 0.07 \, \mathrm{rad\, pF^{-1}}}

\documentclass[%
PRApplied,
nofootinbib,
 amsmath,amssymb,
 reprint,%
]{revtex4-1}

\usepackage{graphicx}
\usepackage{dcolumn}
\usepackage{bm}

\usepackage[utf8]{inputenc}
\usepackage[T1]{fontenc}
\usepackage{mathptmx}

\begin{document}

\preprint{AIP/123-QED}

\title[Non-galvanic calibration and operation of a quantum dot thermometer]{Non-galvanic calibration and operation of a quantum dot thermometer}

\author{J.\,M.\,A.\,Chawner}
 \email{j.chawner@lancaster.ac.uk}
 \affiliation{Department of Physics, Lancaster University, Lancaster, LA1 4YB, United Kingdom}
 
\author{S.\, Barraud}
\affiliation{CEA/LETI-MINATEC, CEA-Grenoble, 38000 Grenoble, France}
\author{M.\,F.\,Gonzalez-Zalba}
\affiliation{Hitachi Cambridge Laboratory, J. J. Thomson Ave., Cambridge, CB3 0HE, United Kingdom}
\thanks{Present address: Quantum Motion Technologies, Windsor House, Cornwall Road, Harrogate, England, HG1 2PW}
\author{S.\,Holt}
\affiliation{Department of Physics, Lancaster University, Lancaster, LA1 4YB, United Kingdom}
\author{E.\,A.\,Laird}
\affiliation{Department of Physics, Lancaster University, Lancaster, LA1 4YB, United Kingdom}
\author{Yu.\,A.\,Pashkin}
\affiliation{Department of Physics, Lancaster University, Lancaster, LA1 4YB, United Kingdom}
\author{J.\,R.\,Prance}
\affiliation{Department of Physics, Lancaster University, Lancaster, LA1 4YB, United Kingdom}

\date{\today}

\begin{abstract}
A cryogenic quantum dot thermometer is calibrated and operated using only a single non-galvanic gate connection. The thermometer is probed with radio-frequency reflectometry and calibrated by fitting a physical model to the phase of the reflected radio-frequency signal taken at temperatures across a small range. Thermometry of the source and drain reservoirs of the dot is then performed by fitting the calibrated physical model to new phase data. The thermometer can operate at the transition between thermally broadened and lifetime broadened regimes, and outside the temperatures used in calibration. Electron thermometry was performed at temperatures between $3.0\,\mathrm{K}$ and $1.0\,\mathrm{K}$, in both a $1\,\mathrm{K}$ cryostat and a dilution refrigerator. In principle, the experimental setup enables fast electron temperature readout with a sensitivity of $\demodsensONE$, at kelvin temperatures. The non-galvanic calibration process gives a readout of physical parameters, such as the quantum dot lever arm. The demodulator used for reflectometry readout is readily available and very affordable.
\end{abstract}

\maketitle

\section{Introduction}

Electron temperature is a fundamental parameter that can limit the performance of low-temperature experiments and applications. Electron thermometry is an essential tool in understating the behaviour of low-temperature circuitry \cite{giazotto2006opportunities}, for example the processors used in quantum computers or devices used to study exotic electronic phases and materials. Accurate and fast electron temperature readout is also a valuable tool for thermodynamic experiments. Increasingly sensitive quantum circuits require delicate and non-invasive electronic thermometry. Quantum dot (QD) and single-electron transistor (SET) conduction thermometry are well established as a powerful approach to monitor electron temperatures \cite{beenakker1991theory, pekola1994thermometry, spietz2003primary, giazotto2006opportunities, maradan2014gaas, koski2015chip, bradley2016nanoelectronic, hahtela2016traceable, iftikhar2016primary, nicoli2019quantum, jones2020progress}. However, these thermometers require the measurement of current through the QD, which can complicate or interfere with other electronic measurements in the experiment. Furthermore, the extra galvanic connections for dot thermometry can be a source of additional noise and parasitic heating. Local charge sensing can be used to probe the occupation of a QD without having to pass a direct current through it \cite{dicarlo2004differential, torresani2013nongalvanic, mavalankar2013non, maradan2014gaas, prance2009electronic}. This approach still requires galvanic connections to read the charge sensor. Similarly, radio frequency (RF) reflectometry techniques have recently allowed QD thermometry to be performed without measuring current through the QD, which is effective outside the lifetime broadened regime \cite{gasparinetti2015fast, ahmed2018primary}. However, for both charge sensing and reflectometry, connections to the source and drain of the dot are still needed to measure the dot lever arm for calibration. 

Here we demonstrate how a completely non-galvanic QD thermometer can be calibrated and used with a single capacitive gate connection, including a calibration of the lever arm with no DC source-drain bias. The QD thermometer has the flexibility to be used on any conducting reservoir and operates in an intermediate regime where the maximum temperature is much lower than the QD charging energy $k_\mathrm{B} T \ll E_\mathrm{c}$ and the minimum temperature is similar to the tunnel coupling $k_\mathrm{B} T \sim \hbar \Gamma$.

\section{Experimental Details}
\begin{figure}
	\centering
	\includegraphics[width=1.0\linewidth]{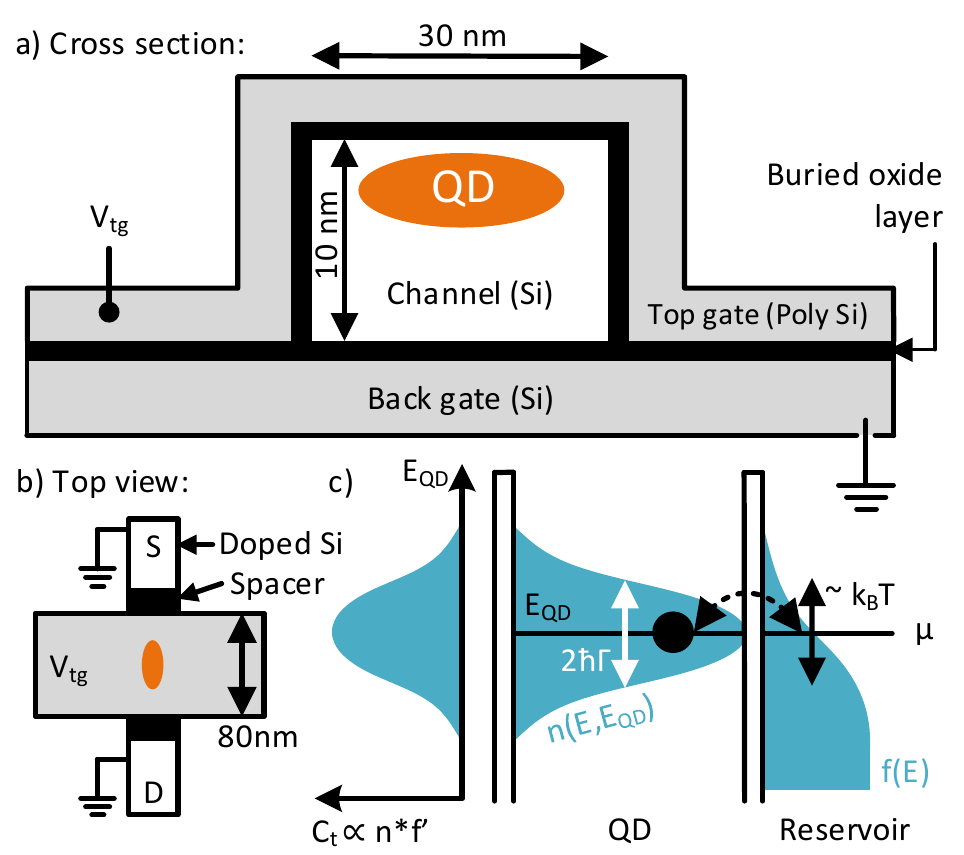}
	\caption{Details of the silicon field-effect transistor QD. \textbf{a)} Cross-section schematic of the device. The transistor consists of an undoped 1-D Si channel, $10\,\mathrm{nm}$ high by $30\,\mathrm{nm}$ wide, with n-doped Si source-drain connections. A polycrystalline silicon top gate, $80\,\mathrm{nm}$ wide, bridges over the channel, separated by a layer of SiO$_2$. A grounded Si back gate is beneath the channel, separated from the channel by $145\,\mathrm{nm}$ thick buried SiO$_2$. \textbf{b)} Top view of device. The source (S) and drain (D) channel connection points are n-doped and behave as a single grounded reservoir during thermometry operation. Two spacers are used to prevent doping of the Si channel under the top gate. \textbf{c)} Energy diagram of the system. The reservoir has an occupation of electron states given by the Fermi-Dirac distribution $f$. The QD energy level $E_\mathrm{QD}$ is broadened from tunnel coupling to the reservoir, with a density of states $n$ described by Eq. (\ref{eq:lor}). The tunnelling capacitance $C_\mathrm{t}$ as a function of $E_\mathrm{QD}$ has a shape proportional to $f'\ast n$.}
	\label{fig:device}
\end{figure}
\begin{figure}
	\centering
	\includegraphics[width=1.0\linewidth]{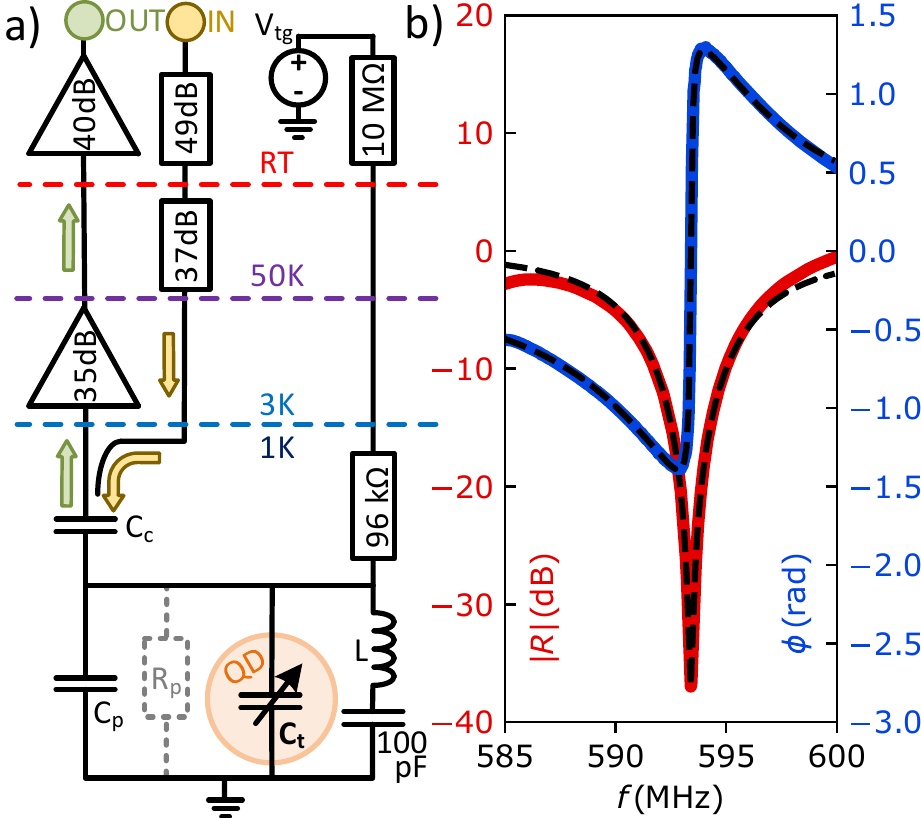}
	\caption{Details of RF circuitry. \textbf{a)} Schematic of the circuit layout. The resonant circuit is comprised of the NbTiN-on-quartz spiral inductor $L$, the parasitic capacitance $C_\mathrm{p}$, coupling capacitance $C_\mathrm{c}$ and the variable QD tunnelling capacitance $C_\mathrm{t}$, which is the physical parameter monitored for thermometry. $C_\mathrm{p}$ includes the geometric gate capacitance $C_\mathrm{tg}$ for modelling purposes. $R_\mathrm{p}$ is a modelled parasitic loss resistance to ground which impacts the resonant circuit Q-factor. The inductor line has a $100 \,\mathrm{pF}$ capacitor to provide a DC break between the top gate and ground. The $96 \,\mathrm{k \Omega}$ resistor limits top gate RF signal loss to the DC line. $V_\mathrm{tg}$ is the controllable DC top gate voltage. IN and OUT represent the RF reflectometry input and output signal, respectively. \textbf{b)} The loaded measurement circuit resonance when coupled to the top gate, with the reflected signal magnitude $|R|$ and phase $\phi$ shown in red and blue, respectively. The black dashed lines show the fitted circuit model, which measures a Q-factor of 63. Thermometry is performed on resonance at $f_0 = 593\,\mathrm{MHz}$, where phase is most responsive.}
	\label{fig:cicuit}
\end{figure}

The QD device used in our experiment is a silicon-on-insulator trigate accumulation-mode field-effect transistor (FET) with a channel length, width and height of $80\,\mathrm{nm}$, $30\,\mathrm{nm}$ and $10\,\mathrm{nm}$ respectively, shown in Fig \ref{fig:device}a and b. For more details about the device, see reference \cite {betz2014high}. At cryogenic temperatures, when a positive sub-threshold DC voltage $V_\mathrm{tg}$ is applied to the top gate, a localisation potential appears in the Si channel underneath the gate. This allows bound electron states to accumulate, and creates a single QD. Corner dots \cite{voisin2014few, ibberson2018electric} do not appear in the structure due to the narrow width of the channel. The FET top gate also acts as a plunger gate for the QD via capacitance $C_\mathrm{tg}$, allowing tuning of the next available energy level in the QD, which is labelled $E_{\mathrm{QD}}$, by adjusting $V_\mathrm{tg}$. The source and drain electrodes are both grounded and act as a single reservoir of electrons with a Fermi level $\mu$ (Fig \ref{fig:device}c). Throughout the experiment, the back gate was grounded. 

The QD energy level $E_\mathrm{QD}$ is broadened by tunnel coupling to the reservoir \cite{kouwenhoven1997electron, ihn2010semiconductor}. This is described with a broadened density of states in the form of a normalised Lorentzian, given by:
\begin{equation}
n(E_{\mathrm{QD}}, E) = \frac{1}{\pi} \frac{\hbar \Gamma}{(E_{\mathrm{QD}} - E)^2 + (\hbar \Gamma)^2}, 
\label{eq:lor}
\end{equation}
where $\hbar$ is the reduced Planck's constant \cite{kouwenhoven1997electron}. The tunnel rate $\Gamma$ between the QD and the reservoir is given by $\Gamma = \Gamma_\mathrm{s} \Gamma_\mathrm{d}/(\Gamma_\mathrm{s}+\Gamma_\mathrm{d})$, where $\Gamma_\mathrm{s}(\mathrm{d})$ is the tunnel rate through source(drain) barrier. The probability $P_{\mathrm{QD}}$ of an excess electron occupying the QD is given by the integral of the product of $n(E_\mathrm{QD}, E)$ and the Fermi-Dirac distribution of electrons in the reservoir $f(E)$ \cite{kouwenhoven1997electron}:
\begin{equation}
P_{\mathrm{QD}}(E_\mathrm{QD}) = \int_{-\infty}^{\infty} f(E)n( E_{\mathrm{QD}}, E) \mathrm{d}E.
\label{eq:prob_int}
\end{equation}
This is the convolution of the two functions, so $P_{\mathrm{QD}}$ becomes: 
\begin{equation}
P_{\mathrm{QD}}(E_\mathrm{QD}) = (f \ast n).
\label{eq:prob_conv}
\end{equation}
Due to the spin degeneracy of the extra electron in the QD, $P_{\mathrm{QD}} = 1/2$ occurs when $E_\mathrm{QD} = \mu \pm k_\mathrm{B}T\mathrm{ln}2$ \cite{hartman2018direct, ahmed2018primary}. The QD has both a constant geometric capacitance and a variable `tunneling' capacitance $C_\mathrm{t}$, given by \cite{esterli2019small,ahmed2018primary,mizuta2017quantum,shevchenko2012multiphoton,shevchenko2012inverse,johansson2006readout}:
\begin{equation}
C_\mathrm{t}(V_\mathrm{tg}) = e \alpha \frac{\partial P_{\mathrm{QD}}}{\partial V_\mathrm{tg}},
\label{eq:ct1}
\end{equation}
where $\alpha = C_\mathrm{tg} / C_\Sigma$ is the top gate lever arm, $C_\Sigma$ is the total QD capacitance, $\partial V_\mathrm{tg} = -\partial E_\mathrm{QD} / e \alpha$, and $e$ is the elementary charge. Inserting Eq. (\ref{eq:prob_conv}) into this definition gives us the tunnelling capacitance of the QD-reservoir system: 
\begin{equation}
C_\mathrm{t}(V_\mathrm{tg})  = e \alpha (f \ast n)' = e \alpha (f' \ast n).
\label{eq:ct2}
\end{equation}
 The derivative of $f$ with respect to $V_\mathrm{tg}$, denoted $f'$, is
\begin{equation}
f'(V_\mathrm{tg}) = \frac{1}{4 k_\mathrm{B} T_\mathrm{e}} \mathrm{cosh}^{-2} \left( \frac{-\alpha e (V_\mathrm{tg} - V_0)}{2 k_\mathrm{B} T_\mathrm{e}}\right),
\label{eq:fermi_dirac_diff}
\end{equation}
where $T_\mathrm{e}$ is the electron temperature of the reservoirs, $V_0$ is the value of $V_\mathrm{tg}$ when $P_{\mathrm{QD}} = 1/2$, and $k_\mathrm{B}$ is the Boltzmann constant. The $T_\mathrm{e}$ dependence of $C_\mathrm{t}$ via Eq. (\ref{eq:fermi_dirac_diff}) is the basis for the non-galvanic QD thermometer. 

The capacitance $C_\mathrm{t}$ was measured by RF reflectometry using the setup shown in Fig \ref{fig:cicuit}a. A resonant circuit was connected to the QD top gate and consisted of a NbTiN-on-quartz spiral inductor $L = 96\,\mathrm{nH}$, and a coupling capacitor $C_\mathrm{c} = 0.18\,\mathrm{pF}$. The inductor is placed parallel to the measured capacitance to help improve the loaded Q-factor, which helps achieve a larger signal for a given change in capacitance \cite{ahmed2018radio}. Modelling the circuit using the measured RF reflection $|R|$ and phase $\phi$ (shown in Fig \ref{fig:cicuit}b) gives a resonance frequency of $f_0 = 593.4 \,\mathrm{MHz}$ with a Q-factor of $63$, a parasitic capacitance of $C_\mathrm{p} = 0.57 \, \mathrm{pF}$ and a parasitic resistance of $R_\mathrm{p} = 43.2 \,\mathrm{k \Omega}$. A modelled circuit loss to ground is represented by resistance $R_\mathrm{p}$, which affects the resonance Q-factor \cite{ahmed2018radio}. The parasitic capacitance $C_\mathrm{p}$ includes the geometric gate capacitance $C_\mathrm{tg}$. The demodulation of $|R|$ and $\phi$ was performed with an `ADL5387' active quadrature demodulator chip. To allow a DC bias $V_\mathrm{tg}$ to be applied to the top gate, a $100 \,\mathrm{pF}$ capacitor was placed after the inductor to avoid DC-shorting the device and to create a good RF ground at the frequency of operation. A $96 \,\mathrm{k \Omega}$ resistor was used to prevent RF signals from escaping via the DC bias line. The resonant frequency depends on total top gate capacitance via $f_0 =  1 /2 \pi \sqrt{L(C_\mathrm{c} + C_\mathrm{p} + C_\mathrm{t})}$. At the circuit resonant frequency, $\phi \propto C_\mathrm{t}$, when $C_\mathrm{t} \ll C_\mathrm{p} + C_\mathrm{c}$ \cite{johansson2006readout, chorley2012measuring, betz2015dispersively, gonzalez2016gate, ahmed2018primary}. This gives a change in reflected signal phase that depends on $T_\mathrm{e}$ according to:
\begin{equation}
\phi - \phi_0 = A e \alpha \left(f'\ast n\right),
\label{eq:phase_conv}
\end{equation}
where $A$ is a fitting constant, which tells us the phase change due to a small change in capacitance, and $\phi_0$ is the circuit phase offset at the resonant frequency when $C_\mathrm{t} \approx 0$. If the constants $A$, $\alpha$ and $\Gamma$ are known, measuring $\phi - \phi_0$ as a function of $V_\mathrm{tg} - V_0$ and fitting the model described by Eq. (\ref{eq:phase_conv}) gives a readout of electron temperature $T_\mathrm{e}$. A demonstration of this technique is shown in Fig \ref{fig:unifit}.

\section{Results}
\begin{figure}
	\centering
	\includegraphics[width=1.0\linewidth]{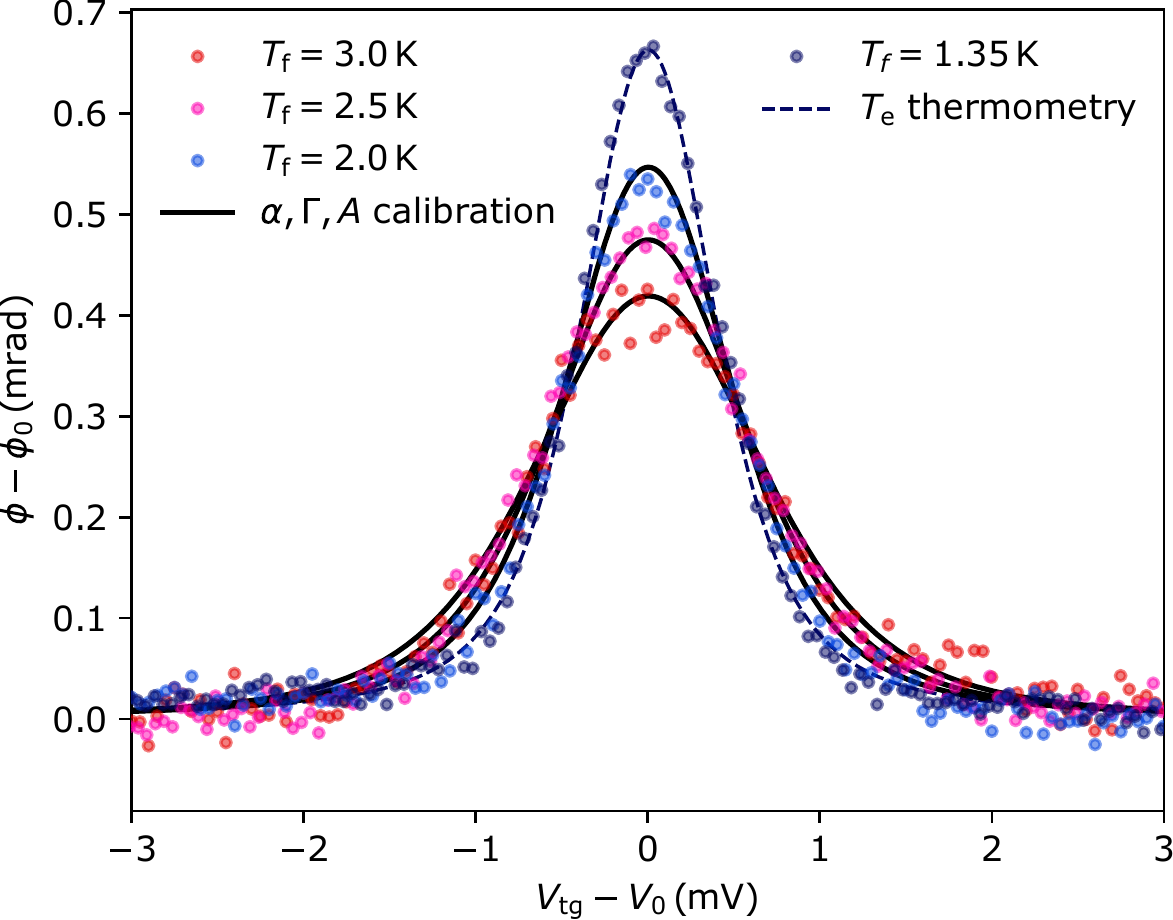}
	\caption{QD thermometer calibration in $1\,\mathrm{K}$ cryostat, showing three experimental phase traces taken at fridge temperatures $T_\mathrm{f} = 2.0,\, 2.5,\, 3.0 \, \mathrm{K}$. Each trace shows the change in reflected signal phase $\phi - \phi_0$ against top gate voltage $V_\mathrm{tg} - V_0$ around the Coulomb peak of the QD. The solid black lines show the least-square fit of Eq. (\ref{eq:phase_conv}) onto the data, assuming the electron temperature $T_\mathrm{e}$ equals the fridge thermometer readout $T_\mathrm{f}$. This calibration procedure estimates $\alpha = \ioleverarm$ , $\Gamma = \iotunnel$ and $A = \ioA $. These three constants are then included within Eq. (\ref{eq:phase_conv}), allowing electron thermometry to be performed, shown here with a $T_\mathrm{e}$ fit to data taken at $T_\mathrm{f} = 1.35 \, \mathrm{K}$, yielding an electron temperature $T_\mathrm{e} = 1.4 \pm 0.1 \,\mathrm{K}$}
	\label{fig:unifit}
\end{figure}
\begin{figure}
	\centering
	\includegraphics[width=1.0\linewidth]{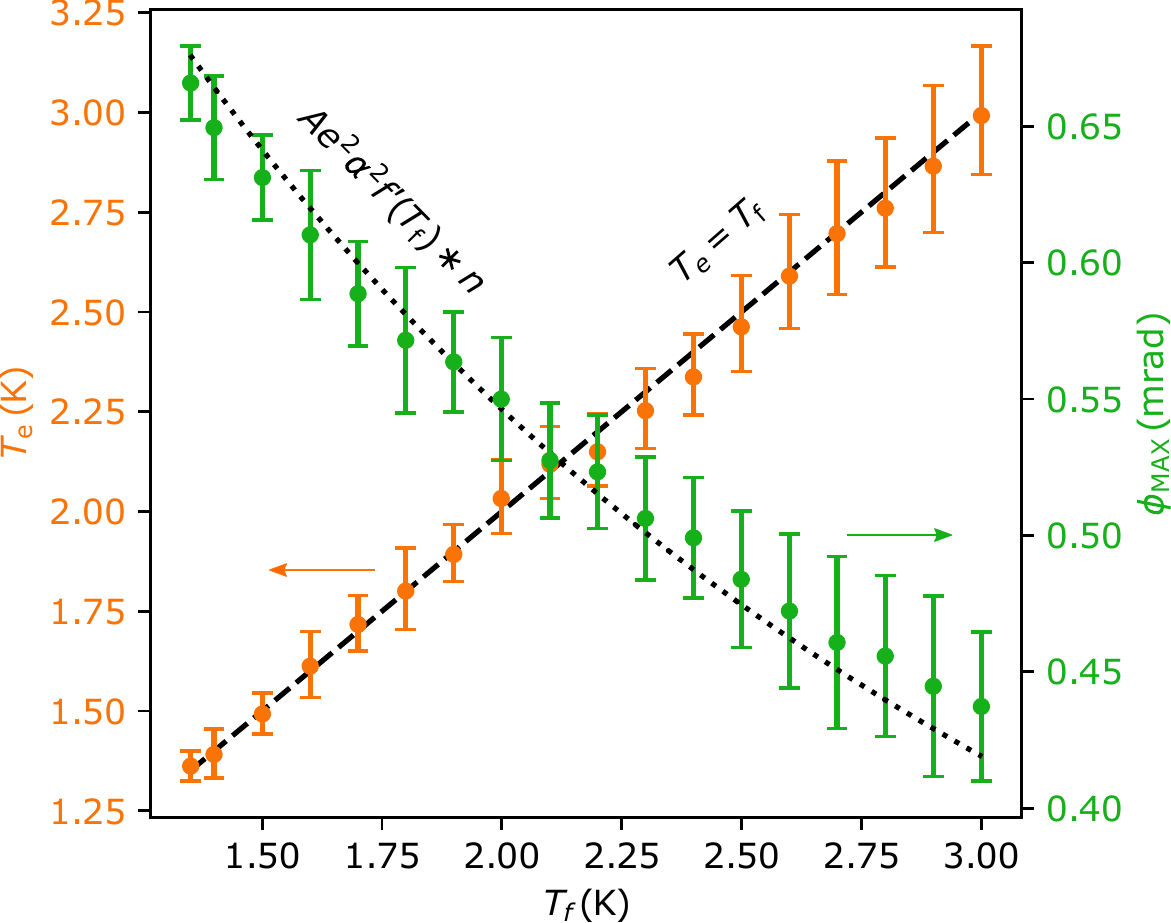}
	\caption{Electron temperature $T_\mathrm{e}$ (orange) and max phase $\phi_{\mathrm{MAX}}$ (green) readout from QD thermometer in $1\,\mathrm{K}$ cryostat. Thermometry readout was generated by fitting $T_\mathrm{e}$, via calibrated Eq. (\ref{eq:phase_conv}), to the phase curve observed by sweeping the $V_\mathrm{tg}$ over the QD Coulomb peak. The peak phase $\phi_{\mathrm{MAX}}$ was measured at $V_{\mathrm{tg}}=V_0$, with no fitting process required. The fridge temperature $T_\mathrm{f}$ was read from a ruthenium oxide fridge thermometer thermally linked to the QD device. The dashed line highlights where $T_\mathrm{e} = T_\mathrm{f}$. The dotted line represents the model prediction from calibrated Eq. (\ref{eq:phase_conv}), assuming $T_\mathrm{e} = T_\mathrm{f}$.}
	\label{fig:thermometry}
\end{figure}
\begin{figure}
	\centering
	\includegraphics[width=1.0\linewidth]{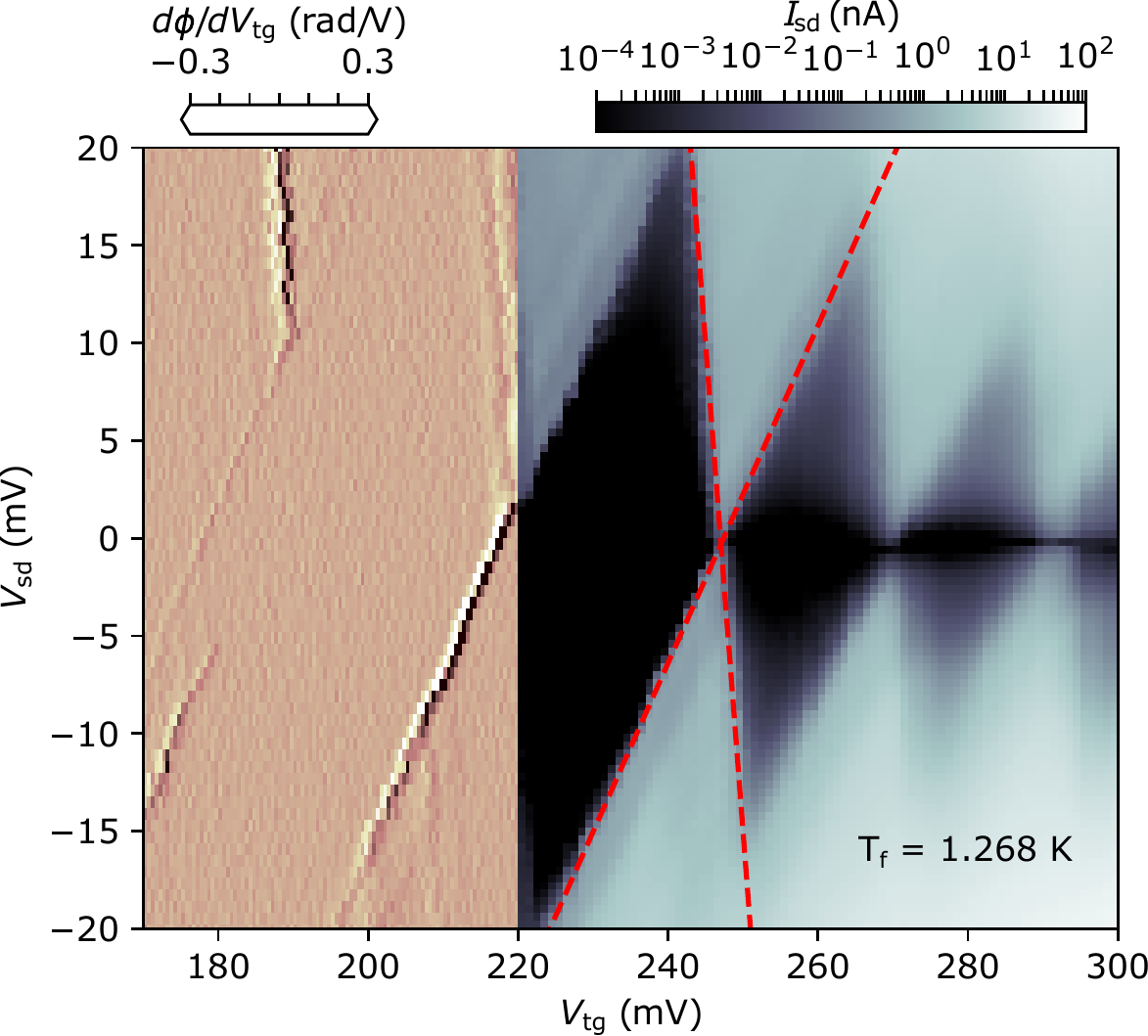}
	\caption{Charge stability diagram measured by reflectometry technique and DC transport in $1\,\mathrm{K}$ cryostat. To the left of $V_\mathrm{tg} = 220\,\mathrm{mV}$, the derivative of reflected signal phase with respect to the top gate voltage $\mathrm{d}\phi/\mathrm{d}V_{\mathrm{tg}}$ is plotted, demonstrating the non-galvanic reflectometry technique. For this measurement, the source and drain connections were grounded. To the right of $V_\mathrm{tg} = 220\,\mathrm{mV}$, the QD source-drain current $|I_{\mathrm{sd}}|$ is plotted in log scale. The fridge temperature was $T_\mathrm{f} = 1.268 \pm 0.001 \, \mathrm{K}$. Red lines highlight the source-drain gradients that match the calibration fit lever arm prediction, $\alpha = \ioleverarm$, via Eq. (\ref{eq:alpha}), crossing at the QD Coulomb peak where the thermometry took place. Here the on-resonance current has a order of magnitude $I_\mathrm{sd} \approx 1-10\,\mathrm{nA}$, which is similar to the single-electron current defined by the calibration fit tunnel rate $e \Gamma \sim \ioisd$.}
	\label{fig:stab_check}
\end{figure}
\begin{figure}
	\centering
	\includegraphics[width=1.0\linewidth]{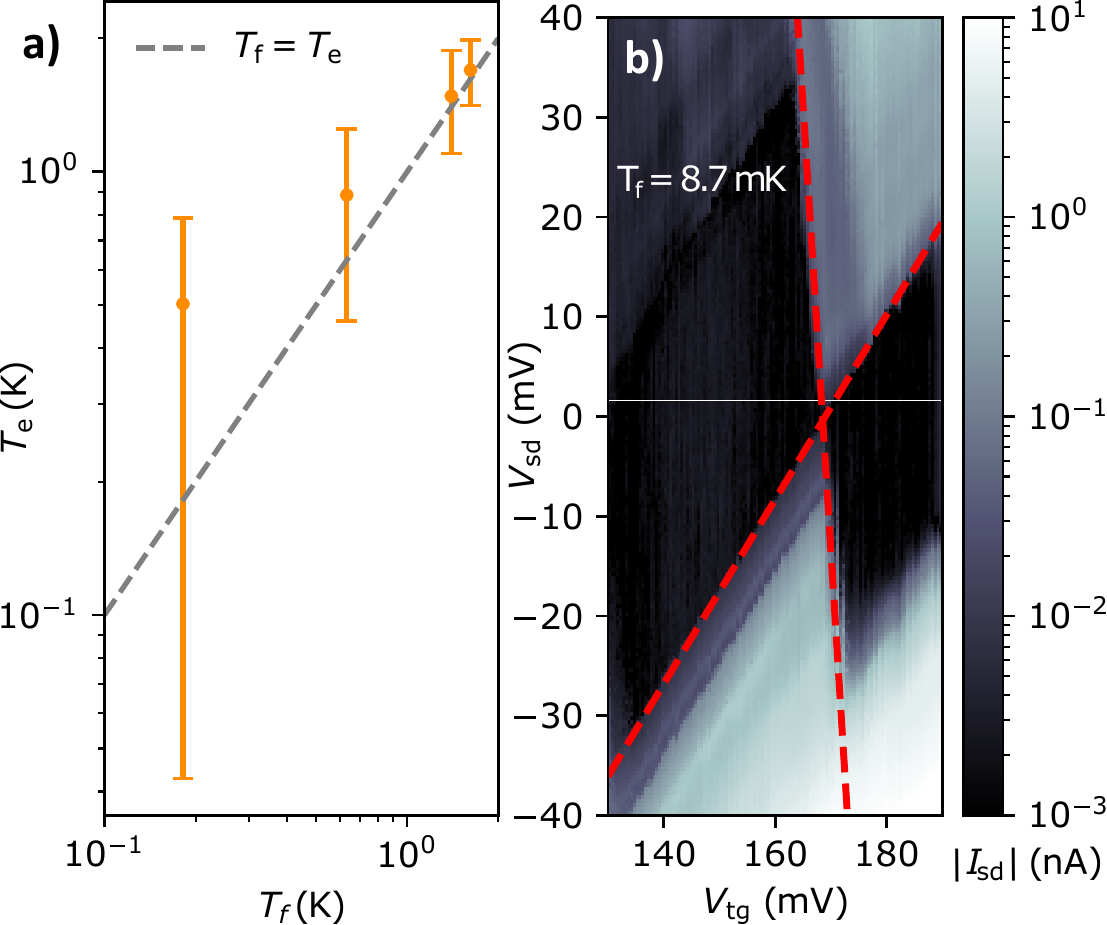}
	\caption{QD Thermometry and stability diagram from dilution refrigerator measurements. \textbf{a)} Re-calibrated QD thermometer readout of electron temperature $T_\mathrm{e}$ compared with fridge thermometer readout $T_\mathrm{f}$, measured near the device within the dilution refrigerator. The QD and fridge thermometers start to disagree below $1\,\mathrm{K}$, but still agree within one standard deviation of confidence (For details on how the error is determined, see section II in the Supplemental Material \cite{supplemetal}). \textbf{b)} Source-drain current stability diagram of QD mounted in dilution refrigerator.  The temperature of the fridge was held at $T_\mathrm{f} = 8.70 \pm 0.05 \, \mathrm{mK}$.  Red lines highlight the source-drain gradients that match the calibration fit lever arm prediction, $\alpha = \bfleverarm$, via Eq. (\ref{eq:alpha}), crossing at the QD Coulomb peak where the thermometry took place.}
	\label{fig:BF_stab_et}
\end{figure}
The QD thermometer was measured in two fridge systems: a cryogen-free $1\,\mathrm{K}$ cryostat\footnote{Oxford Instruments `Io'} and a cryogen-free dilution refrigerator\footnote{BlueFors `LD250'}, where it was calibrated and operated. In both systems, the fridge temperature was monitored by a ruthenium oxide resistance fridge thermometer\footnote{Model ref `ROTH-GEN' in the 1K cryostat and `RuO2.RX-102B' in the dilution refrigerator} mounted alongside the QD thermometer during data collection. The reading of the ruthenium oxide fridge thermometer is denoted as $T_\mathrm{f}$.

In the $1\,\mathrm{K}$ cryostat, calibration was done by reading the reflected signal phase across a number of sweeps of $V_\mathrm{tg}$ at a set of fridge temperatures $T_\mathrm{f} = 2.0, \, 2.5, \, 3.0 \, \mathrm{K}$. A single least-squares fit using Eq. (\ref{eq:phase_conv}) was performed on the collective phase traces for this set of temperatures (Fig \ref{fig:unifit}). It was assumed that the electron temperature is well thermalised with the fridge temperature at $2 \,\mathrm{K}$ and above, and so during the calibration the condition $T_\mathrm{e} = T_\mathrm{f}$ was applied to Eq. (\ref{eq:phase_conv}). For each phase trace, $\phi_0$ and $V_0$ were individually fitted. The calibration fit produced an estimate for the values $\alpha = \ioleverarm$ , $\Gamma = \iotunnel$ and $A = \ioA$. This implies $\Gamma \gg 2\pi f_0$, therefore dissipative components were neglected and the cyclic tunnelling was considered adiabatic. With these three constants defined, the thermometer was calibrated and ready for operation. 

To use the QD thermometer, $\phi - \phi_0$ was measured as a function of $V_\mathrm{tg} - V_0$ and fitted with the calibrated Eq. (\ref{eq:phase_conv}) to give a readout of electron temperature $T_\mathrm{e}$. A series of thermometry readings were taken at varying fridge temperatures between $3.0\,\mathrm{K}$ and $1.3\,\mathrm{K}$. Fridge temperature $T_\mathrm{f}$ was monitored for each QD thermometer reading of $T_\mathrm{e}$ (Fig \ref{fig:thermometry}). The QD thermometer agreed with the fridge thermometer across the range of temperatures, even at temperatures below the calibration range. It is worth noting that this works in the intermediate regime where $k_\mathrm{B} T \sim \hbar \Gamma$ because the tunnel broadening is taken into account within the physical model. For quicker electron temperature readout, the QD can be tuned to where $V_\mathrm{tg}=V_0$ so that $\phi - \phi_0 = \phi_\mathrm{MAX}$, which was directly converted to a electron temperature via Eq. (\ref{eq:phase_conv}), using the previously calibrated values of $\alpha$, $\Gamma$ and $A$. This approach worked effectively, even at fridge temperatures below the calibration data.  

Finally, to provide an independent confirmation of the calibration process, the QD source and drain connections were ungrounded to apply a source-drain voltage $V_\mathrm{sd}$ across the QD and measure a charge stability diagram of the device. Using the relationship
\begin{equation}
\alpha = \frac{1}{m_\mathrm{d} - m_\mathrm{s}}, \label{eq:alpha}
\end{equation}
where $m_\mathrm{d} = \Delta V_{\mathrm{tg}}^\mathrm{d} / \Delta V_{\mathrm{sd}}^\mathrm{d}$ is the gradient along the the ‘drain resonance’ side of a Coulomb diamond and $m_\mathrm{s}= \Delta V_{\mathrm{tg}}^\mathrm{s} / \Delta V_{\mathrm{sd}}^\mathrm{s}$ is the ‘source resonance’ side, we can see the lever arm $\alpha = \ioleverarm$ matches well with the Coulomb diamond geometry in Fig \ref{fig:stab_check}. This demonstrates that the lever arm of a QD can be obtained using one non-galvanic gate connection and measurements spanning a range of fridge temperatures. The order of magnitude of source-drain current $|I_\mathrm{sd}|$ from the unblockaded QD was found to be in the order of $\sim 10\,\mathrm{nA}$. This agrees with the calibration fit of the total tunnel rate constant $\iotunnel$, which equates to a source-drain current of $e \Gamma = \ioisd$ for a single electron transport channel. 

To study operation at lower temperatures, the QD thermometer was mounted into a dilution refrigerator. The calibration fit was performed as described above, with phase data taken at $1300 \, \mathrm{mK}$ and $1600 \, \mathrm{mK}$. Within this system, the three calibration constants were found to be $\alpha = \bfleverarm$, $\Gamma = \bftunnel$ and $ A = \bfA$. The change in $A$ is attributed to the differences between the two fridge systems affecting the RF electronics, such as parasitic capacitance $C_\mathrm{p}$ changing due to a different metallic geometry near to the QD chip. Both $\alpha$ and $\Gamma$ are sensitive to the shape and position of the QD in the Si channel, which are likely to be different after a thermal cycle of the device.

The electron temperature readout $T_\mathrm{e}$ from the QD thermometer in the dilution refrigerator agreed with the fridge temperature readout $T_\mathrm{f}$ above $1\,\mathrm{K}$, despite the fact that $ k_\mathrm{B} T_\mathrm{f} < \hbar\Gamma $ (Fig \ref{fig:BF_stab_et}a). Below $1\,\mathrm{K}$ there was deviation of $T_\mathrm{e}$ away from $T_\mathrm{f}$, although the point $T_\mathrm{e} = T_\mathrm{f}$ remains within one standard deviation of experimental uncertainty (For details on how the error is determined, see section II in the Supplemental Material \cite{supplemetal}). The deviation of $T_\mathrm{e}$ readout and its increase in uncertainty occurs because the QD energy level is strongly tunnel-broadened and the response of the phase trace to temperature becomes weaker (For details on the influence of tunnel broadening, see section III and FIG. 4 in the Supplemental Material \cite{supplemetal}). With a reduced $\Gamma$ the QD thermometer would work at lower temperatures. This can be achieved by adjusting the device design geometry. Checking the charge stability diagram, the predicted lever arm $\alpha = \bfleverarm$ matches the Coulomb diamond geometry well (Fig \ref{fig:BF_stab_et}b). The dilution refrigerator experiment has an average white noise phase sensitivity of $\demodsensPHA$, in this case dominated by the measurement chain. With this equipment and the $\phi_\mathrm{MAX}$ measurements demonstrated in Fig \ref{fig:thermometry}, the QD thermometer setup could achieve a potential sensitivity of $\demodsensTHREE$ at $3.0 \, \mathrm{K}$, and $\demodsensONE$ at $1.3 \, \mathrm{K}$ (For details on the thermometer sensitivity, see section I in the Supplemental Material \cite{supplemetal}).

\section{Conclusions}

We have described the successful calibration and operation of a QD thermometer readout via a single capacitive connection in two separate cryostats,  which introduces a new level of simplicity and versatility for measuring electron temperature. The calibration uses limited data to generate a physical model of the QD-reservoir system, which correctly estimates physical parameters such as the QD lever arm. Electron thermometry was successfully performed with the calibrated QD thermometer in a $1.0\,\mathrm{K}$ to $3.0\,\mathrm{K}$ range. The QD thermometer was also used for faster readout by monitoring the phase when the QD has an occupation probability of $1/2$. In this mode of operation, the noise floor of the measurement should allow for a sensitivity of $\demodsensONE$ and $\demodsensTHREE$, at $1.3 \, \mathrm{K}$ and $3.0 \, \mathrm{K}$ respectively. This process worked with the same QD chip in both cryostats. The QD thermometer can operate even in the case where $k_\mathrm{B}T_\mathrm{e} < \hbar \Gamma$, however with the system and techniques used here, the thermometry uncertainty starts to increase below $1\,\mathrm{K}$ due to strong tunnel coupling overriding the temperature dependence. Careful analysis of the thermometry uncertainty reveals the coldest limit of the QD thermometer, when the electron temperature readout confidence boundary increases beyond a usable size. A redesign of the QD device to reduce the tunnel rate would be needed to decrease the uncertainty at lower temperatures. The ability to fully calibrate and operate a non-galvanic electron thermometer with a single RF line simplifies the application of the device substantially. This device provides a versatile, sensitive and effective tool for monitoring electron temperature in nanoelectronic devices at cryogenic temperatures.

\begin{acknowledgments}
We thank Xiao Collins, Kunal Lulla Ramrakhiyani, Michael Thompson and Alex Jones for their assistance.

This work was funded and supported by the European Microkelvin Platform (the European Union’s Horizon 2020 research and innovation programme, Grant Agreement No. 824109), the UK EPSRC (EP/N019199/1), the ERC (818751) and Hitachi Europe Ltd.

The data that support the findings of this study are available at https://doi.org/10.17635/lancaster/researchdata/403, including descriptions of the data sets.
\end{acknowledgments}


%

\end{document}


\title{Supplementary Material for `Non-galvanic calibration and operation of a quantum dot thermometer'}


\author{J.\,M.\,A.\,Chawner}
\email{j.chawner@lancaster.ac.uk}
\affiliation{Department of Physics, Lancaster University, Lancaster, LA1 4YB, United Kingdom}

\author{S.\, Barraud}
\affiliation{CEA/LETI-MINATEC, CEA-Grenoble, 38000 Grenoble, France}
\author{M.\,F.\,Gonzalez-Zalba}
\affiliation{Hitachi Cambridge Laboratory, J. J. Thomson Ave., Cambridge, CB3 0HE, United Kingdom}
\thanks{Present address: Quantum Motion Technologies, Windsor House, Cornwall Road, Harrogate, England, HG1 2PW}

\author{S.\,Holt}
\affiliation{Department of Physics, Lancaster University, Lancaster, LA1 4YB, United Kingdom}
\author{E.\,A.\,Laird}
\affiliation{Department of Physics, Lancaster University, Lancaster, LA1 4YB, United Kingdom}
\author{Yu.\,A.\,Pashkin}
\affiliation{Department of Physics, Lancaster University, Lancaster, LA1 4YB, United Kingdom}
\author{J.\,R.\,Prance}
\affiliation{Department of Physics, Lancaster University, Lancaster, LA1 4YB, United Kingdom}

\date{\today}

\maketitle

\section{Determination of QD Thermometer sensitivity}

\begin{figure}
	\centering
	\includegraphics[width=\linewidth]{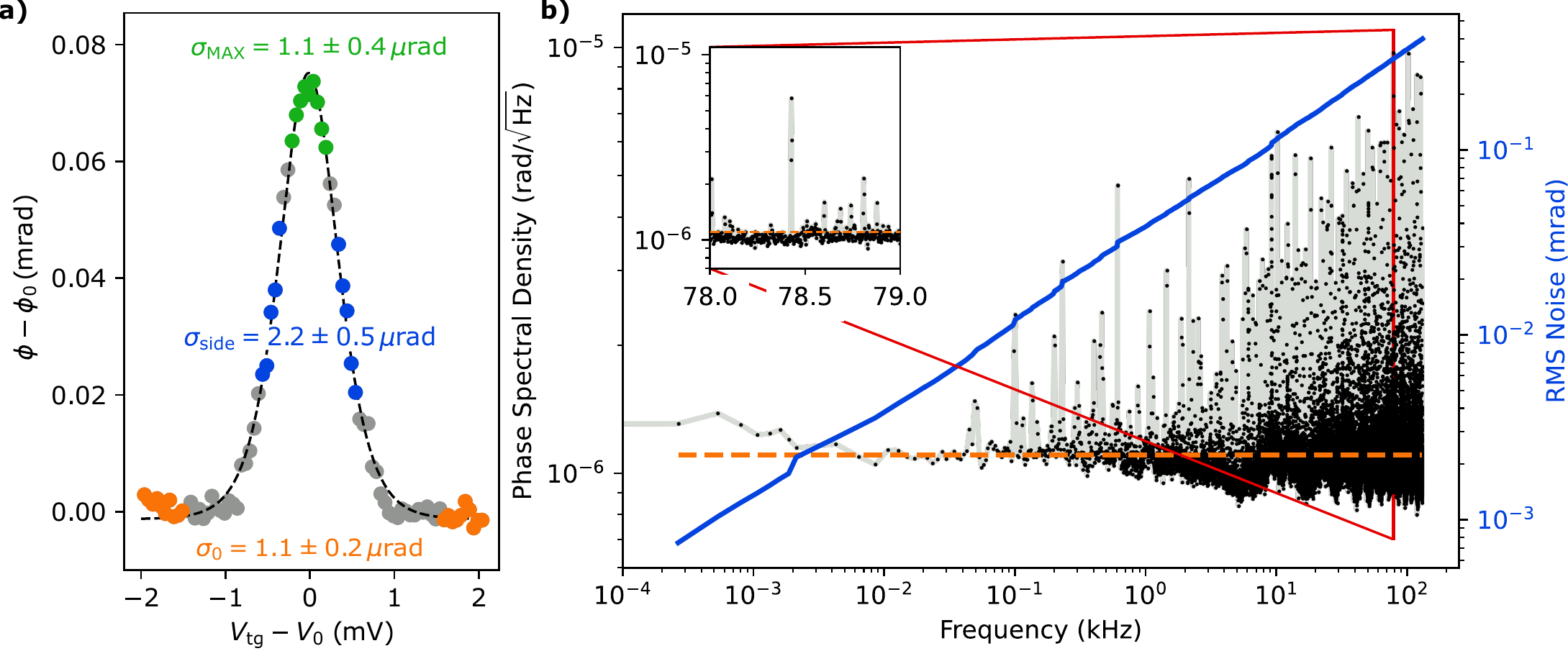}
	\caption{Reflected phase modulation and noise measured in the dilution refrigerator. \textbf{a)} A phase trace taken at $T_\mathrm{f} = 0.18\,\mathrm{K}$. Each data point was measured for 3 seconds. The RMS phase noise for readings taken at the tails, $\sigma_0$ (orange), sides, $\sigma_\mathrm{side}$ (blue), and peak, $\sigma_\mathrm{MAX}$ (green), of the trace are equivalent, i.e $\sigma_0 \approx \sigma_\mathrm{side} \approx \sigma_\mathrm{MAX}$. The black dashed line shows a fit to the data that is used to estimate the RMS deviation of each point. \textbf{b)} An RF signal was reflected off the QD thermometer at $T_\mathrm{f} = 8.7 \,\mathrm{mK}$, and a ADL5387 quadrature demodulator chip was used to extract the phase from the reflected signal. (This is detailed in the paper within the text and Figure 2a). The black points, connected with grey lines, show the spectral density of the readout. The orange dashed line corresponds to the average spectral density over the whole bandwidth, which is equal to a effective white noise phase sensitivity of $\demodsensPHA$. The inset displays a zoomed in section of the spectral density, which shows that the noise peaks seen at higher frequencies are extremely narrow compared to their typical spacing. The blue line shows the total RMS phase noise, which is the integral of the spectral density between $0.27\,\mathrm{Hz}$ and the frequency indicated.}
	\label{fig:phase_spectra}
\end{figure}
The sensitivity of the thermometer is determined by its response to temperature and the noise in the measurement of the phase of the reflected RF signal. Within the dilution refrigerator, a phase trace was measured at $T_\mathrm{f} = 0.18\,\mathrm{K}$, shown in FIG \ref{fig:phase_spectra}a). The RMS phase noise is consistent across the Coulomb peak, so it is assumed that the noise arises principally from the measurement chain and is not intrinsic to the device. It is presumed that any noise intrinsic to the QD, such as charge noise, lies somewhere underneath the measured noise floor. The reflected signal phase $\phi$ from the blockaded quantum dot (QD) was recorded at a sample rate of $2.6 \times 10^5 \,\mathrm{s}^{-1}$ for $0.5 \,\mathrm{s}$. This was done $128$ times and averaged, producing the spectral density in FIG \ref{fig:phase_spectra}b). The result is generally flat, with narrow spikes of noise at higher frequencies. These spikes are much narrower than their typical spacing (most only having a single data point), therefore contribute relatively little to the total noise. In principle, they could be easily avoided with the use of a lock-in measurement technique. For example, a Zurich Instruments UHFLI could perform the demodulation and subsequently lock-in to a component of the demodulated signal. Averaging the spectral density over the entire bandwidth gives an effective white noise phase sensitivity of $\demodsensPHA$. Using $\mathrm{d} \phi / \mathrm{d} T_\mathrm{e}$ from Eq. (7) in the paper text, with the condition $V_{\mathrm{tg}}=V_0$, the phase spectrum can be converted to temperature spectrum. The best case phase reponsivity, $|\mathrm{d} \phi / \mathrm{d} T_\mathrm{e}| = 0.10 \pm 0.01 \,\mathrm{mrad/K}$ at $3.0 \, \mathrm{K}$ and $|\mathrm{d} \phi / \mathrm{d} T_\mathrm{e}| = 0.24 \pm 0.02 \,\mathrm{mrad/K}$ at $1.3 \, \mathrm{K}$ (from FIG 4 in the paper), gives a corresponding effective temperature sensitivity of $\demodsensTHREE$ at $3.0 \, \mathrm{K}$, and $\demodsensONE$ at $1.3 \, \mathrm{K}$.

\section{Electron Thermometry Uncertainty}

\begin{figure}
	\centering
	\includegraphics[scale=0.9]{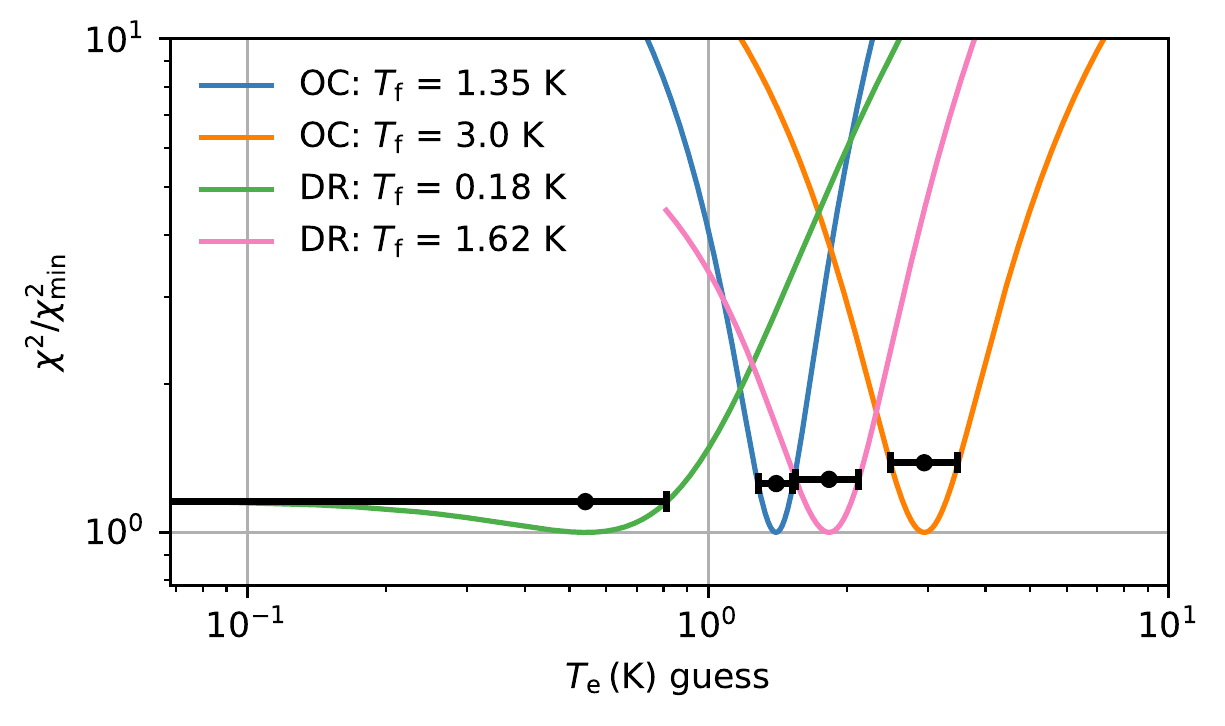}
	\caption{Goodness of fit $\chi^2 / \chi_\mathrm{min}^2$ against electron temperature $T_\mathrm{e}$ guess for a variety of phase traces. Two phase traces are from measurements taken within a cryogen-free, $1\,\mathrm{K}$ cryostat, `OC', and were measured at $T_\mathrm{f} = 3.0 \,\mathrm{K}$ and $1.35 \,\mathrm{K}$. Two more phase traces are from dilution refrigerator, `DR', measurements taken at $T_\mathrm{f} = 1.6 \,\mathrm{K}$ and $0.2 \,\mathrm{K}$. The black points represent the $T_\mathrm{e}$ fit where $\chi^2$ is minimum. The height of these points is the confidence boundary $\chi^2 = \chi_\mathrm{min}^2 + 1$ which defines the confidence region shown by the error bars.}
	\label{fig:fit_error}
\end{figure}
The QD thermometer calibration process described in the paper text gives an estimation of $\alpha$, $\Gamma$ and $A$. These three constants are included in the phase model (described by Eq. (7) in the paper text), leaving only electron temperature $T_\mathrm{e}$ as a fitting parameter, along with $\phi_0$ and $V_0$ which are trivial to define for each phase trace. For a given phase trace, each guess of $T_e$ returns a goodness of fit that can be quantified with $\chi^2$ via
\begin{equation}
\chi^2 = \sum_{i = 0}^{N -1} \frac{[\phi^i - \phi(V_\mathrm{tg}^i | T_\mathrm{e}\mathrm{,}\, \phi_0\mathrm{,}\, V_0)]^2}{\sigma (N -3)},
\end{equation}
where $N$ is the number of data points in the phase trace, $\phi^i$ and $V_\mathrm{tg}^i$ are the values of phase and top gate voltage respectively for the $i^\mathrm{th}$ data point \cite{press2007numerical}. The standard deviation of phase readout for each data point, $\sigma$, is constant over the whole phase trace. However, $\sigma$ does vary for each trace, for example if the temperature is different. The minimum value of $\chi^2$, which we will label $\chi^2_\mathrm{min}$, corresponds to the final fit estimate for $T_\mathrm{e}$, and is always $\chi^2_\mathrm{min} < 10$. To work out the certainty of this measurement, a constant confidence boundary of $\chi^2_\mathrm{min} + 1$ surrounds $\chi^2_\mathrm{min}$, creating an interval which represents one standard deviation of confidence. FIG \ref{fig:fit_error} demonstrates this process with a few example phase traces; $T_\mathrm{f} = 3.0 \,\mathrm{K}$ and $1.35 \,\mathrm{K}$ from the cryogen-free $1\,\mathrm{K}$ cryostat, $T_\mathrm{f} = 1.6 \,\mathrm{K}$ and $0.1 \,\mathrm{K}$ from the dilution refrigerator. At temperatures below $1\,\mathrm{K}$ the confidence interval starts to broaden, and can extend to $T_\mathrm{e} = 0$. At this point it can be said the QD thermometer has reached its lower temperature limit for operation. An explanation for the broadening of the confidence region is given in Section \ref{sec:energy}.
 
\section{Energy Scales}
\label{sec:energy}

\begin{figure}
	\centering
	\includegraphics[width = \linewidth]{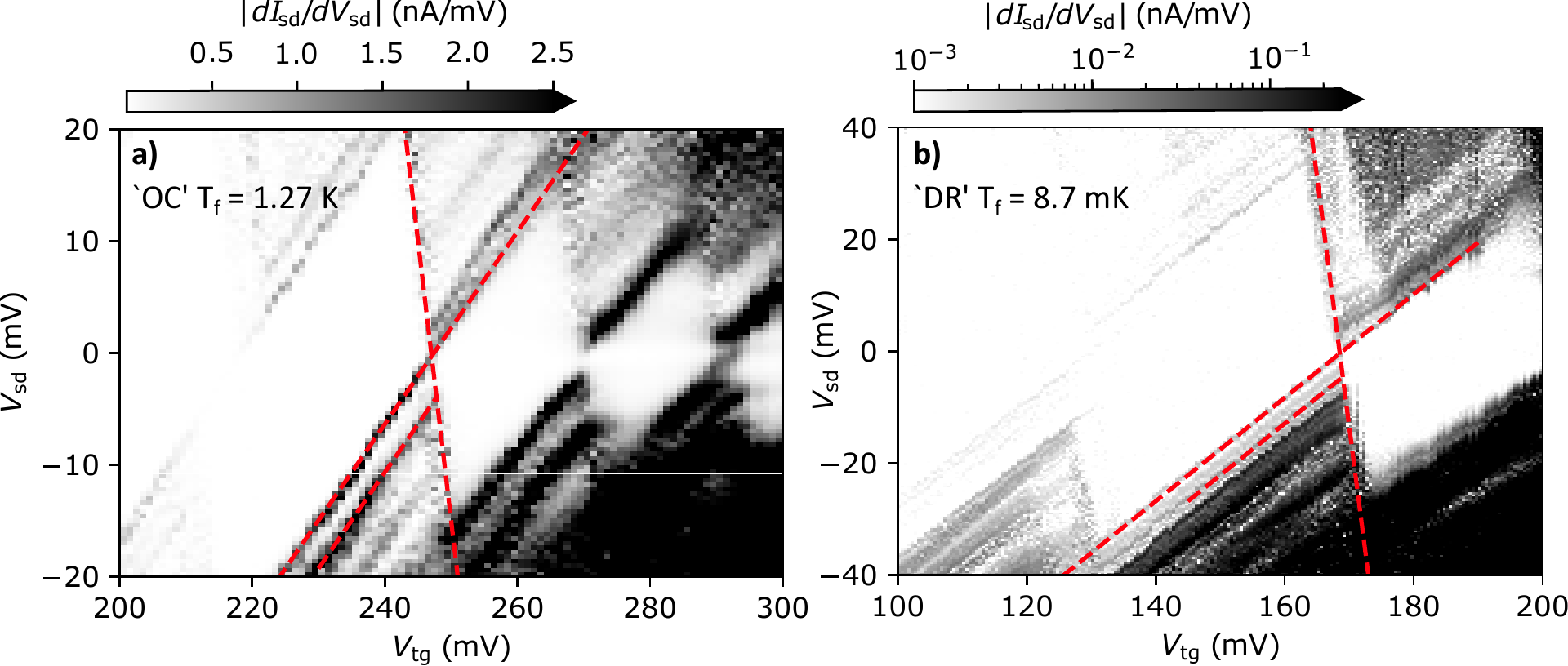}
	\caption{Stability diagrams for QD in two fridge systems: \textbf{a)} a cryogen-free 1K cryostat `OC' and \textbf{b)} a dilution refrigerator `DR'. $V_\mathrm{tg}$ is the top gate voltage, $V_\mathrm{sd}$ is source-drain voltage and $I_\mathrm{sd}$ is source-drain current. For each diagram, the crossed reds line highlight the Coulomb resonance where the thermometry was performed, and the gradients correspond to the lever arm as measured from the QD calibration process, $\alpha = \ioleverarm$ and $\bfleverarm$ for OC and DR, respectively. The lowest red line highlights the first excited state of the QD.}
	\label{fig:stab}
\end{figure}
\begin{table}
	\centering
	\begin{tabular}{|c|c|c|c|c|}
		\hline 
		\textbf{Energy Scale}$\,\mathrm{(meV)}$ & $k_\mathrm{B} T_\mathrm{e}$ & $\hbar \Gamma$ & $E_\mathrm{con}$ & $E_\mathrm{c}$  \\ 
		\hline 
		Cryogen-free 1K cryostat & \thermalenio & \tunnelenio & \excitedenio & \chargingenio \\
		\hline
		Dilution refrigerator & \thermalenbf & \tunnelenbf & \excitedenbf & \chargingenbf \\
		\hline
		
	\end{tabular} 
	\caption{Energy scales associated with the experiments in the two fridge systems. $T_\mathrm{e}$ is the electron temperature. The experiments took place over a range of temperatures, so the thermal energy scale is shown as a range. $\Gamma$ is the QD tunnel rate which defines the tunnel coupling. $E_\mathrm{con}$ is the confinement energy for the first excited state in the QD. $E_\mathrm{c}$ is the charging energy for the QD.}
	\label{tab:energy_scales}
\end{table}
Several energy scales are involved in the experiments that impact the behaviour of the QD. The tunnel coupling $\hbar \Gamma$ is measured from the QD thermometer calibration process. The QD excited state confinement energy $E_\mathrm{con}$ and QD charging energy $E_\mathrm{c} = e \Delta V_\mathrm{sd}$ can be observed in FIG \ref{fig:stab} for both the 1K cryostat and dilution refrigerator experiments. To extract the voltage $\Delta V_\mathrm{sd}$, the average height was taken from the Coulomb diamonds on either side of the conduction resonance where the thermometry was performed. The thermal energy $k_\mathrm{B} T$ is extracted from the range of temperatures used in each fridge during the experiments. All the energy scales for both systems are stated in Table \ref{tab:energy_scales} for comparison. For the cryogen-free 1K cryostat experiment, the QD energy condition is
\begin{equation}
k_\mathrm{B} T_\mathrm{e} \sim \hbar \Gamma < E_\mathrm{con} < E_\mathrm{c},
\end{equation}
so the Coulomb-blockaded QD is affected but not dominated by tunnel coupling. For the dilution refrigerator experiment the energy condition is
\begin{equation}
k_\mathrm{B} T_\mathrm{e} < \hbar \Gamma  < E_\mathrm{con} < E_\mathrm{c},
\end{equation}
so the Coulomb-blockaded QD is strongly tunnel-broadened and the tunnelling capacitance has much weaker temperature dependence.

\begin{figure}
	\centering
	\includegraphics[scale=0.9]{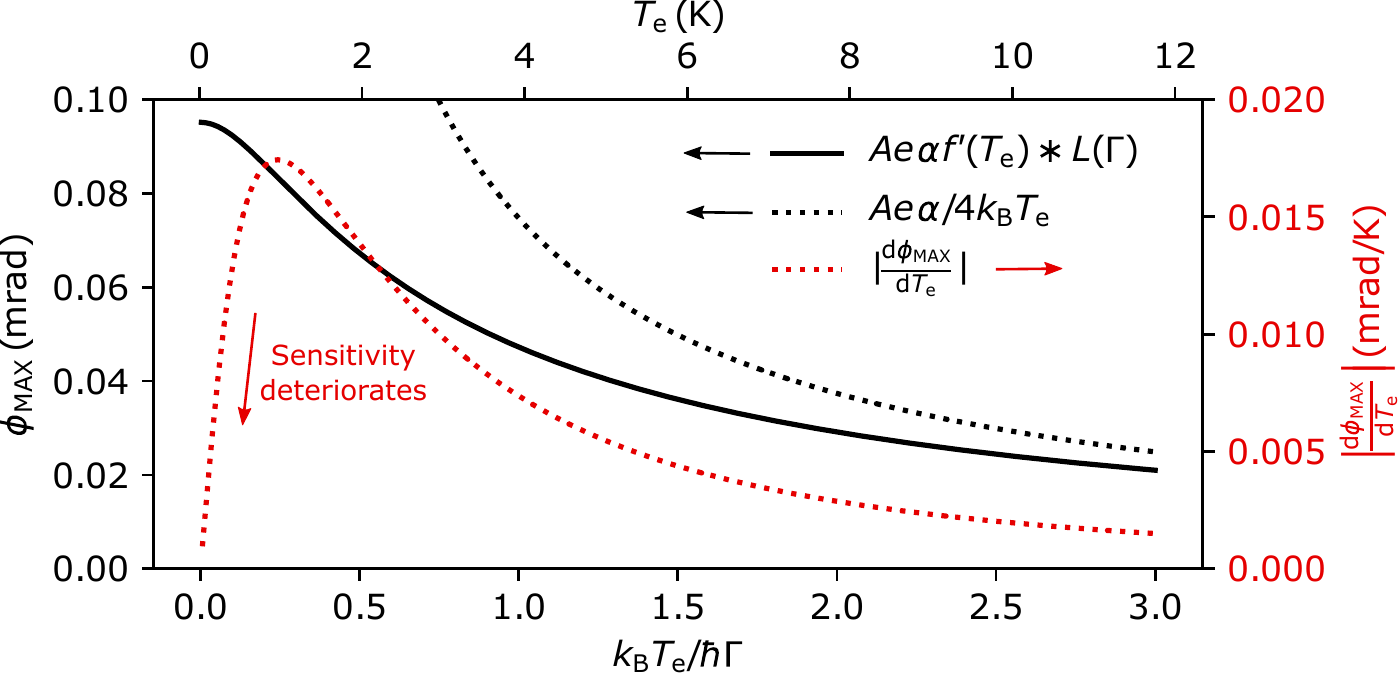}
	\caption{Predicted maximum phase $\phi_\mathrm{MAX}$ (solid black line) against electron temperature, when $V_\mathrm{tg} = V_0$. The upper x-axis shows the electron temperature $T_\mathrm{e}$ and the lower x-axis shows the corresponding ratio between the thermal energy and tunnel coupling $k_\mathrm{B} T_\mathrm{e}/\hbar \Gamma$. The derivative of $\phi_\mathrm{MAX}$ with respect to $T_\mathrm{e}$ is shown with a red dotted line. The prediction is generated from the thermometer calibration in the dilution refrigerator experiment detailed in section III of the paper text. The dotted black line shows the predicted $\phi_\mathrm{MAX}$ assuming there is no tunnel coupling, i.e $\Gamma = 0$, which will converge with the tunnel-coupled model at high temperature. Below $T_\mathrm{e} = 1\,\mathrm{K}$ the maximum phase $\phi_\mathrm{MAX}$ begins to plateau, which indicates the loss of temperature sensitivity, and therefore the lower operational limit of the electron temperature.}
	\label{fig:pmax}
\end{figure}
In the case of the dilution refrigerator experiment, detailed in section III of the paper, the $T_\mathrm{e}$ readout uncertainty increases as the temperature is reduced below $\sim 1 \,\mathrm{K}$. This occurs since the thermal energy scale is much smaller relative to the tunnel broadening, and so the phase trace shape has much weaker temperature dependence. This can be seen in FIG \ref{fig:pmax}, which uses the predicted maximum phase of the phase trace, $\phi_\mathrm{MAX}$, as an indicator of the thermometer's temperature sensitivity. The prediction is based off the calibration of the QD thermoemter in the dilution refrigerator experiment. At high temperatures, the tunnel coupling does not impact the thermometer phase significantly from the theoretical case of zero tunnel coupling. As the temperature is reduced to $k_\mathrm{B} T_\mathrm{e}/\hbar \Gamma \approx 2$, the maximum phase deviates significantly away from the prediction with negligible tunnel coupling. The sensitivity deteriorates at even lower temperatures where $k_\mathrm{B} T_\mathrm{e}/\hbar \Gamma \approx 1/4$, which in this particular case is $T_\mathrm{e} \approx 1\,\mathrm{K}$. Approaching $0\,\mathrm{K}$, the phase trace converges on a Lorentzian shape defined by the tunnel broadening, and so the maximum phase plateaus, losing its temperature sensitivity. This defines the lower limit at which the QD thermometer can operate.

\begin{figure}
	\centering
	\includegraphics[scale=0.9]{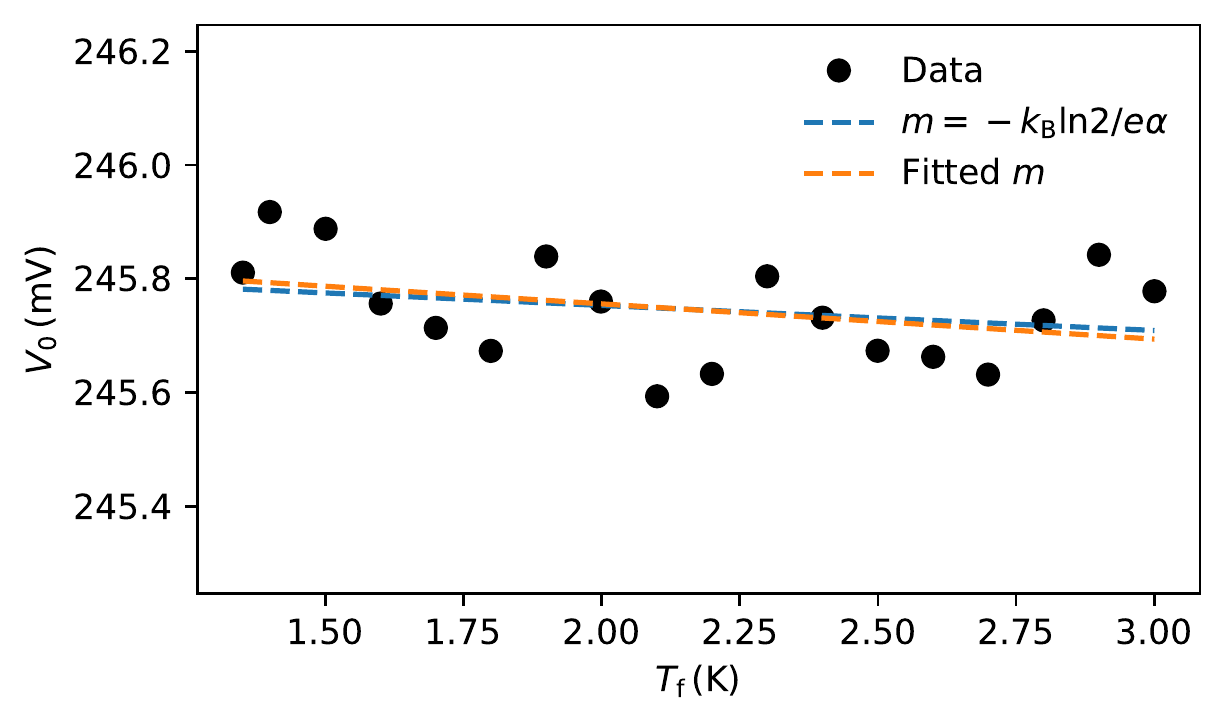}
	\caption{Location of the top gate voltage $V_0$ where $P_\mathrm{QD} = 1/2$, against fridge temperature $T_\mathrm{f}$ in the cryogen-free 1K cryostat. $m$ equals the gradient $\mathrm{d}V_0 / \mathrm{d}T_\mathrm{f}$. The blue line shows the theoretical temperature dependence of $V_0$ with a gradient of $m = k_\mathrm{B} \mathrm{ln}2 / e \alpha\, \mathrm{VK}^{-1}$. The orange line shows a linear fit to the data which predicts a gradient of $m = - (6 \pm 4) \times 10^{-5} \, \mathrm{VK}^{-1}$. }
	\label{fig:charge_degeneracy_loc}
\end{figure}
It has been shown that for a QD-reservoir system, the imbalance of spin degeneracy between an extra electron in the QD and the absence of an extra electron creates a temperature dependence for the QD energy with $1/2$ occupation probability, described by $E_\mathrm{QD} = \mu \pm k_\mathrm{B}T\mathrm{ln}2$ \cite{hartman2018direct, ahmed2018primary}. The top gate voltage value for $P_\mathrm{QD} = 1/2$, labelled $V_0$, is recorded for each phase trace measured across a range of temperatures. FIG \ref{fig:charge_degeneracy_loc} shows a subtle negative correlation between $V_0$ and the fridge temperature $T_\mathrm{f}$. The theoretical gradient $\mathrm{d}V_0 / \mathrm{d}T_\mathrm{f} = -k_\mathrm{B} \mathrm{ln}2 / e \alpha\, \mathrm{VK}^{-1} = -4.4 \times 10^{-5} \, \mathrm{VK}^{-1}$ is certainly believable when compared with the measured $V_0$, and a linear fit returns $\mathrm{d}V_0 / \mathrm{d}T_\mathrm{f} = - (6 \pm 4) \times 10^{-5} \, \mathrm{VK}^{-1}$. Although the correlation is clouded by charge noise, the relationship between $V_0$ and $T_\mathrm{f}$ is consistent with the theory.


%